# Estimating potential infection transmission routes in hospital wards using wearable proximity sensors

P. Vanhems[1,2], A. Barrat[3,4,5], C. Cattuto[5], J.-F. Pinton[6], N. Khanafer[1,2], C. Régis[2], B. Kim[7], B. Comte[7], N. Voirin[1,§]

[1]Hospices Civils de Lyon, Hôpital Edouard Herriot, Service d'Hygiène, Epidémiologie et Prévention, Lyon, France
[2]Université de Lyon; université Lyon 1; CNRS UMR 5558, laboratoire de Biométrie et de Biologie Evolutive, Equipe Epidémiologie et Santé Publique, Lyon, France
[3]Aix Marseille Université, CNRS, CPT, UMR 7332, 13288 Marseille, France
[4]Université de Toulon, CNRS, CPT, UMR 7332, 83957 La Garde, France
[5]Data Science Lab, ISI Foundation, Torino, Italy
[6]Laboratoire de Physique de l'Ecole Normale Supérieure de Lyon, CNRS UMR 5672, Lyon, France
[7]Hospices Civils de Lyon, Hôpital Edouard Herriot, Service de gériatrie, Lyon, France
[§]Corresponding author

# Abstract

**Background**

Contacts between patients, patients and health care workers (HCWs) and among HCWs represent one of the important routes of transmission of hospital-acquired infections (HAI). A detailed description and quantification of contacts in hospitals provides key information for HAIs epidemiology and for the design and validation of control measures.

**Methods and Findings**

We used wearable sensors to detect close-range interactions ("contacts") between individuals in the geriatric unit of a university hospital. Contact events were measured with a spatial resolution of about 1.5 meters and a temporal resolution of 20 seconds. The study included 46 HCWs and 29 patients and lasted for 4 days and 4 nights. 14,037 contacts were recorded overall, 94.1% of which during daytime. The number and duration of contacts varied between mornings, afternoons and nights, and contact matrices describing the mixing patterns between HCW and patients were built for each time period. Contact patterns were qualitatively similar from one day to the next. 38% of the contacts occurred between pairs of HCWs and 6 HCWs accounted for 42% of all the contacts including at least one patient, suggesting a population of individuals who could potentially act as super-spreaders.

**Conclusions**

Wearable sensors represent a novel tool for the measurement of contact patterns in hospitals. The collected data can provide information on important aspects that impact the spreading patterns of infectious diseases, such as the strong heterogeneity of contact numbers and durations across individuals, the variability in the number of contacts during a day, and the fraction of repeated contacts across days. This variability is however associated with a marked statistical stability of contact and mixing patterns across days. Our results highlight the need for such measurement efforts in order to correctly inform mathematical models of HAIs and use them to inform the design and evaluation of prevention strategies.

**Introduction**

The control of hospital-acquired infections (HAI) is largely based on preventive procedures derived from the best available knowledge of potential transmission routes. The accurate description of contact patterns between individuals is crucial to this end, as it can help to understand the possible transmission dynamics and the design principles for appropriate control measures. In particular, the mutual exposures between patients and health-care workers (HCWs) have been documented for bacterial and viral transmission since decades [1,2,3]. Transmission might be the result of effective contact, as in the cases of *S. aureus* [4,5], *K. pneumoniae* [6] or rotavirus [7], of exposure to contaminated aerosols, as for *M. tuberculosis* [8], or the result of exposure to droplets, as for influenza [9]. Some pathogens such as influenza can also be transmitted by different routes. Although close-range proximity and contacts between individuals are strong determinants for potential transmissions, obtaining reliable data on these behaviors remains a challenge [10].

Data on contacts between individuals in specific settings or in the general population are most often obtained from diaries and surveys [11,12,13,14] and from time-use records [15]. These approaches provide essential information to describe contacts patterns and inform models of infectious disease spread. The gathered data, however, often lack the longitudinal dimension [10,12,16] and the high spatial and temporal resolution needed to accurately characterize the interactions among individuals in specific environments such as hospitals. Moreover, they are subject to potential biases due to behavioral modifications due to the presence of observers, to short periods of observation, and especially to missing information and recall biases. Evaluating biases and understanding the accuracy of the collected data is therefore a difficult task [16].

In this context, the use of electronic devices has recently emerged as an interesting complement to more traditional methods [10]. In particular, wearable sensors based on active Radio-Frequency IDentification (RFID) technology have been used to measure face-to-face proximity relations between individuals with a high spatio-temporal resolution in various contexts [17] that include social gatherings [18,19], schools [20,21] and hospitals [22,23]. The amount of available data, however, is still very limited, high-resolution contact data relevant for the epidemiology of infectious diseases are scarce, and the longitudinal aspects of contact patterns have not been investigated in detail, prompting further investigation.

In this paper we report on the use of wearable proximity sensors [17] to measure the numbers and durations of contacts between individuals in an acute care geriatric unit of a university hospital. We investigate the variability of contact patterns as a function of time, as well as the differences in contact patterns between individuals with different roles in the ward. We document the presence of individuals with a high number of contacts, who could be considered as potential super-spreaders of infections. Some implications of our results regarding prevention and control of hospital-acquired infections are discussed.

**Materials and Methods**

*Study design and data collection*

The measurement system, developed by the SocioPatterns collaboration [24], is based on small active RFID devices ("tags") that are embedded in unobtrusive wearable badges and exchange ultra-low-power radio packets [17,18,21,23]. The power level is tuned so that devices can exchange packets only when located within 1–1.5 meters of one another, i.e., package exchange is used as a proxy for distance (the tags do not directly measure distances). Individuals were asked to wear the devices on their chests using lanyards, ensuring that the RFID devices of two individuals can only exchange radio packets when the persons are facing each other, as the human body acts as a RF shield at the frequency used for communication. In summary the system is tuned so that it detects and records close-range encounters during which a communicable disease infection could be transmitted, for example, by cough, sneeze or hand contact. The information on face-to-face proximity events detected by the wearable sensors is relayed to radio receivers installed throughout the hospital ward (bedrooms, offices and hall).

The system was tuned so that whenever two individuals wearing the RFID tags were in face-to-face proximity the probability to detect such a proximity event over an time interval of 20 seconds was larger than 99%. We therefore define two individuals to be in "contact" during a 20-second interval if and only if their sensors exchanged at least one packet during that interval. A contact is therefore symmetric by definition, and in case of contacts involving three or more individuals in the same 20-second interval, all the contact pairs were considered. After the contact is established, it is considered ongoing as long as the devices continue to exchange at least one packet for every subsequent 20s interval. Conversely, a contact is considered broken if a 20-second interval elapses with no exchange of packets. We emphasize that this is an operational definition of the human proximity behavior that we choose to quantify, and that all the results we present correspond to this precise and specific definition of "contact". We make the raw data we collected available to the public as Supplementary Information datasets S1-S5 and on the website of the SocioPatterns collaboration (www.sociopatterns.org).

Data were collected in a short stay geriatric unit (19 beds) of a university hospital of almost 1000 beds [3] in Lyon, France, from Monday, December 6, 2010 at 1:00 pm to Friday, December 10, 2010 at 2:00 pm. During that time, 50 professional staff worked in the unit and 31 patients were admitted. We collected data on the contacts between 46 staff members (92% participation rate) and 29 patients (94% participation rate). The participating staff members were 27 nurses or nurses' aides, 11 medical doctors and 8 administrative staff.

In the ward, all rooms but 2 were single-bed rooms. Each day 2 teams of 2 nurses and 3 nurses' aides worked in the ward: one of the teams was present from 7:00 am to 1:30 pm and the other from 1:30 pm to 8:00 pm. An additional nurse and an additional nurse' aid were moreover present from 9:00 am to 5:00 pm. Two nurses were present during the nights from 8:00 pm to 7:00 am. In addition, a physiotherapist and a nutritionist were present each day at various points in time, with no fixed schedule, and a social counselor and a physical therapist visited on demand (in out analysis they are considered as nurses). Two physicians and 2 interns were present from 8:00 am to 17:00 pm each day. Visits were allowed from 12:00 am to 8:00 pm but visitors were not included in the study.

*Ethics and privacy*

In advance of the study, staff members and patients were informed on the details and aims of the study. A signed informed consent was obtained for each participating patient and staff member. All participants were given an RFID tag and asked to wear it properly at all times. No personal information was associated with the tag: only the professional category of each HCW and the age of the patients were collected. The study was approved by the French national bodies responsible for ethics and privacy, the "Commission Nationale de l'Informatique et des Libertés" (CNIL, http://www.cnil.fr) and the "Comité de Protection des personnes" (http://www.cppsudest2.com/) of the hospital.

*Data analysis*

Individuals were categorized in four classes according to their activity in the ward: patients (PAT), medical doctors (physicians and interns, MED), paramedical staff (nurses and nurses' aides, NUR) and administrative staff (ADM). MED and NUR professionals form a group named HCW.

The contact patterns were analyzed using both the numbers and the durations of contacts between individuals. For each individual we measured the number of other distinct individuals with whom she/he had been in contact, as well as the total number of contact events she/he was involved in, and the total time spent in contact with other individuals. These quantities were aggregated for each class and for each pair of role classes in order to define contact matrices that describe the mixing patterns between classes of individuals.

The longitudinal evolution of the contact patterns was studied by considering, in addition to the entire study duration, several shorter time intervals: We divided the study duration into 5 daytime periods (7:00 am to 8:00 pm) and 4 nights (8:00 pm to 7:00 am); daytime periods were divided in morning (7:00 am to 1:30 pm) and afternoon (1:30 pm to 8:00 pm) shifts, and we also considered data aggregated on a 1-hour timescale.

We finally considered the similarity of contact patterns between successive days, by measuring the fraction of contacts that were repeated from one day to the next, as such information is particularly relevant when modeling spreading phenomena [18,25,26].

**Results**
*Number of contacts*

Overall, 14,037 contacts occurred during the study, with a cumulative duration of 648,480s (approx. 10,808 minutes or 180 hours). 10,616 contacts (75.6%) included at least one NUR, 4,003 (28.5%) included at least one MED, and 3,849 (27.4%) at least one patient. Table 1 reports the average number and duration of contacts of individuals in each class over the whole study duration. Most contacts involve at least one NUR and/or one MED, and NURs and MEDs have on average the largest number of contacts, as well as the largest cumulative duration in contact. Large standard deviations are however observed: the distributions of the contact durations and of the numbers and cumulative durations of contacts are broad, as also observed in many other contexts [20,21,23,27]. Important variations are observed even within each role class. In particular, contacts of much larger duration than the average are observed with a non-negligible frequency.

*Contacts between classes of individuals*

The total number of contacts between individuals belonging to specific classes is reported in Table 2 and the corresponding contact matrices are shown in Figure 1. We report contact matrices giving the total numbers and cumulative durations of contacts between individuals of given classes, as well as contact matrices taking into account the different numbers of individuals in each class. Contacts were most frequent between two NURs (5,310 contacts, 38%), followed by NUR-PAR contacts (2,951 contacts, 21%), and by contacts between two MEDs (2,136 contacts, 15%). Very few contacts between PATs or between members of the ADM group were observed.

*Longitudinal study*

As reported in Table 3, among the 14,037 contacts detected, 13,206 (94.1%) occurred during daytime, for a total duration of 612,900s (approx. 10,215 min or 170 h). 831 contacts (5.9%) occurred during nights (lasting 35,580s, approx. 593 min or 10 h). On average we recorded 2,265 contacts per morning, 1,041 per afternoon, and 207 per night.

The evolution of the number of contacts at the more detailed resolution of one-hour time windows is reported in Figure 2. The number of contacts varied strongly over the course of a day, but the evolution was similar from one day to another (for day 1 and day 5, contacts were recorded after 1:00 pm and before 2:00 pm respectively, see Methods), with very few contacts at night and a maximum around 10-12 am. The number of contacts between individuals of specific classes also depends on the period of the day. Contacts between NURs, and between NURs and PATs, were predominant in the morning while contacts between MEDs remained similar between mornings and afternoons. Overall, 63.3% of contacts between NURs and PATs occurred on the morning, 25.5% on the afternoon and 9.2% during the night.

Figure 3 reports the contact matrices giving the numbers of contacts between individuals of specific classes for each morning, afternoon and night. The absolute numbers of contacts varied from one morning (resp., afternoon or night) to the next, but the mixing patterns remained very similar. Differences were observed between morning, afternoon and night patterns. The main difference between morning and afternoon periods came from larger numbers of contacts involving NURs in the morning. Almost only contacts involving NURs and PATs were observed at night.

Although the aggregated observables reported above are very similar from day to day, the precise structure of the daily contact network varied strongly: the fraction of common neighbors of an individual between two different days is on average just of 48%. This value is smaller than the one observed in a school [21], but much larger than the one measured for the attendees of a conference [18].

"Super-contactors" among HCWs (NURs and MEDs)

The cumulative number and duration of contacts of each individual, as identified by his/her badge identification number, are reported in Figure 4 and Table 4. A small number of HCWs accounted for most of the contacts observed between HCWs and PATs, both in terms of number and cumulative duration. For instance, 6 NURs (representing 16% of all HCWs) accounted for 42.1% of the number of contacts and 44.3% of the cumulated duration of the contacts with PATs (number of contacts and cumulative duration of contacts of a given individual are strongly correlated, $r=0.98$). The number of distinct individuals contacted by a given individual was also correlated with the total number of contacts of the same individual ($r=0.69$). These 6 HCWs had a much larger number and duration of contacts than average, as shown in Table 4.

**Discussion**

The objective of the present study was to describe in detail the contacts between individuals in a healthcare setting. Such data can help to accurately inform computational models of the propagation of infectious diseases and, as a consequence, to improve the design and implementation of prevention or control measures based on the frequency and duration of contacts.

Numbers and duration of contacts were characterized for each class of individuals and for individuals belonging to given class pairs, yielding contact matrices that represent important inputs for realistic computational models of nosocomial infections. As also measured in other contexts [17,20,21,23,27], the numbers and durations of contacts display large variations even across individuals of the same class: the resulting distributions were broad, with no characteristic time scale. As a consequence, even though the average durations of contacts were rather short, contacts of much longer durations than average occured with non-negligible frequency.

Contacts involving either two NURs or between NURs and PATs accounted for the majority of contacts, both in terms of numbers and of global durations. Very few contacts occurred between PATs: this might be a specificity of wards with mostly single rooms, and other wards in which patients are not alone in a room or in which they move around more might yield more numerous contacts between PATs. These results are consistent with previous studies [23,28] carried out in pediatrics, surgery and intensive care units, and provide additional evidence that nurses and assistants may be the most essential target group for prevention measures [27,28].

The detailed information about the number and duration of contacts also allowed us to highlight the presence of a limited number of "super-contactors" among HCWs who account for a large part of all contacts. A large number of contacts could correspond to different situations, namely to contacts with many different patients, or to many contacts with few patients. Our results show that the cumulated number of contacts and the number of distinct persons contacted are correlated; this indicates that in the hospital context under study the super-contactors have contacts with many different patients. They could therefore potentially play the role of super-spreaders, whose importance in the spread of infectious agents has been highlighted both theoretically [29,30] and empirically [31]. This suggests that their role class should be targeted for prevention measures.

These results are in concordance with the central role of HCWs in hospital wards, as repeated contacts with patients are often necessary for the quality of healthcare. However, since outbreaks of measles and influenza involving this population have been observed [32,33], the possibility for HCWs to be super-contactors emphasizes the need to reduce their exposure to infection and to limit the risk of transmission to patients. This should stimulate the strict implementation of preventive measures including hand washing, vaccination, or wearing of masks [34]. In addition, HCWs could be warned against the risk brought forth by unnecessary large numbers or long durations of contacts, especially with patients.

Limiting the contacts of HCWs (either with PATs or with other HCWs) might however not be feasible without altering the quality of care. In this respect, the investigation of the temporal evolution of the numbers of contacts may help envision and discuss changes in the organization of care during epidemic or pandemic periods. The numbers of contacts varied indeed greatly along the course of each day, clearly highlighting the periods of the day (here, the mornings) during which transmission could occur with higher probability. The high numbers of contacts during mornings may indicate a potential overexposure to infection for PATs and NURs, and one may imagine a different organization toward a smoothing of the number of contacts throughout mornings and afternoons. This would decrease the density of contacts, in particular between NURs, at each specific moment, while maintaining the daily number and duration of contacts between NURs and PATs, and overall tend to limit their overexposure [35]. The potential efficacy of such or other changes in the healthcare organization should of course be tested through numerical simulations of spreading phenomena, and their feasibility would moreover need to be asserted through discussions with the staff.

The measurement of contact patterns by means of wearable sensors presents strengths and limitations that are worth discussing. Strong advantages are the versatility of the sensing strategy (i.e., the unobtrusiveness of wearable sensors and the prompt deployability of receivers) and the fact that it does neither require the constant presence of external observers nor interfere with the delivery of care in the ward. Another strength lies in the high spatial and temporal resolution: behavioral differences across role classes can be detected, and longitudinal studies are possible. High participation ratios are also crucial: similarly to a previous study in another hospital [23], the rate of acceptance among HCWs and patients turned out to be very high (92%). The information meetings held before the study, providing a clear exposition of the scientific objectives and of the privacy aspects, most probably played an important role in achieving such a high participation rate.

The versatility of a system based on wearable sensors and easily deployable data receivers makes it possible to repeat similar studies in different environments and to compare results across contexts [19]. In particular, several of the reported findings are very similar to those described in [23] in a different hospital, situated in a different country, and in a different type of ward (paediatric): large variability in the cumulative duration of contacts, small number of contacts between patients, and large numbers and durations of contacts between NURs. Repeating measurements in the same ward and in other wards represents an important step towards understanding the similarities and differences of contact patterns in hospital settings, and allows to generalize the observations to more correctly inform models.

The measurement approach we used here has also several limitations. Contacts were defined as face-to-face proximity, without any information on physical contact between individuals. Therefore, the assumption that the number of contacts reflects disease exposure can be appropriate for respiratory infections such as influenza, or for similar diseases that can be transmitted by various routes at a distance of 1 meter around an index case [36]. The use of close-range proximity as a proxy for the transmission of bacterial infection acquired by cross transmission, such as *S aureus* or *Enterobacteriacea,* is more questionable. Other factors related to specific attributes of individuals (e.g., vaccination or immunosuppression), of the microbial agent (e.g., resistance or virulence) or of the environment (e.g., specialty of ward) may also alter the relationship between contact frequency/duration and transmission. In this respect, a validation with simultaneous direct observation and human annotation of the contacts would be of particular interest.

Finally, it is difficult to assess whether individuals modified their behavior in response to wearing RFID badges, but direct observation indicated that HCWs were focusing on their daily activities and most probably were not influenced by the presence of the badges. Badges were not proposed to visitors and this potential external source of infection was not studied.

*Conclusions and future work*

This study complements previous work [22,23,27,28,30,34] and provides data that can be used to explore the spread of infection in confined settings through mathematical and computational modeling. Models of transmission within hospitals might be based on contact matrices such as those presented here, and used to better understand the epidemiology of different types of microbial agents, to assess the impact of control measures, and to help improve the delivery of care during emergency epidemic situations. In our study, specific mixing patterns were observed between different classes of individuals, showing a clear departure from homogeneous mixing, as it is expected in a hospital setting, and highlighting the relevance of correctly informed contact matrices. Moreover, although an important turnover between the persons in contact with a given individual was observed across different days, and although the average contact durations between different classes of individuals varied between mornings, afternoons and nights, the contact patterns remained statistically very similar across successive days. These results suggest that, in order to correctly inform computational models, data collected over just a few hours might be insufficient, but that measures lasting 48 hours would be sufficient to evaluate the statistical properties of contact patterns as well as the mixing patterns between individual classes, and to estimate the similarity between the contacts of an individual across days. The statistical features of the gathered data could then be used to model contact patterns over longer time scales.

The scarcity of contact data [10,37] calls for further measurement campaigns to validate and consolidate the results across other hospital units, other contexts, and over longer periods of time. Additional data sets would also be useful to build and test proxies that could replace systematic detailed measurement of contact patterns, such as the ones put forward in [15,38,39].

In order to explore the relationship between complex contacts network and the spreading of infections, it would be particularly interesting to collect simultaneously high-resolution contact data and microbiological data describing the infection status of participating individuals. Combining these heterogeneous sources of information within appropriate statistical models would allow elucidating the relation between the risk of disease transmission and contacts patterns, in order to disentangle transmission likelihood from contact frequency. Finally, feedback of the results to HCWs could be an innovative pedagogical tool in health care settings.

**Acknowledgments**

We are particularly grateful to all patients and the hospital staff who volunteered to participate in the data collection.

**Author Contributions**

Conceived and designed the experiments: PV AB CC JFP NK CR BK BC NV. Performed the experiments: AB CC JFP NK CR NV. Analyzed the data: PV AB NV. Wrote the paper: PV AB CC NV.

**References**

1. Albrich WC, Harbarth S (2008) Health-care workers: source, vector, or victim of MRSA? Lancet Infect Dis 8: 289-301.

2. Linquist JA, Rosaia CM, Riemer B, Heckman K, Alvarez F (2002) Tuberculosis exposure of patients and staff in an outpatient hemodialysis unit. Am J Infect Control 30: 307-310.

3. Vanhems P, Voirin N, Roche S, Escuret V, Regis C, et al. (2011) Risk of influenza-like illness in an acute health care setting during community influenza epidemics in 2004-2005, 2005-2006, and 2006-2007: a prospective study. Arch Intern Med 171: 151-157.

4. Chamchod F, Ruan S (2012) Modeling the spread of methicillin-resistant Staphylococcus aureus in nursing homes for elderly. PLoS One 7: e29757.

5. Orendi JM, Coetzee N, Ellington MJ, Boakes E, Cookson BD, et al. (2010) Community and nosocomial transmission of Panton-Valentine leucocidin-positive community-associated meticillin-resistant Staphylococcus aureus: implications for healthcare. J Hosp Infect 75: 258-264.

6. Richards C, Alonso-Echanove J, Caicedo Y, Jarvis WR (2004) Klebsiella pneumoniae bloodstream infections among neonates in a high-risk nursery in Cali, Colombia. Infect Control Hosp Epidemiol 25: 221-225.

7. Trop Skaza A, Beskovnik L, Zohar Cretnik T (2011) Outbreak of rotavirus gastroenteritis in a nursing home, Slovenia, December 2010. Euro Surveill 16.

8. Shenoi SV, Escombe AR, Friedland G (2010) Transmission of drug-susceptible and drug-resistant tuberculosis and the critical importance of airborne infection control in the era of HIV infection and highly active antiretroviral therapy rollouts. Clin Infect Dis 50 Suppl 3: S231-237.

9. Baker MG, Thornley CN, Mills C, Roberts S, Perera S, et al. (2010) Transmission of pandemic A/H1N1 2009 influenza on passenger aircraft: retrospective cohort study. BMJ 340: c2424.

10. Read JM, Edmunds WJ, Riley S, Lessler J, Cummings DA (2012) Close encounters of the infectious kind: methods to measure social mixing behaviour. Epidemiol Infect 140: 2117-2130.

11. Beutels P, Shkedy Z, Aerts M, Van Damme P (2006) Social mixing patterns for transmission models of close contact infections: exploring self-evaluation and diary-based data collection through a web-based interface. Epidemiol Infect 134: 1158-1166.

12. McCaw JM, Forbes K, Nathan PM, Pattison PE, Robins GL, et al. (2010) Comparison of three methods for ascertainment of contact information relevant to respiratory pathogen transmission in encounter networks. BMC Infect Dis 10: 166.

13. Mikolajczyk RT, Akmatov MK, Rastin S, Kretzschmar M (2008) Social contacts of school children and the transmission of respiratory-spread pathogens. Epidemiol Infect 136: 813-822.

14. Mossong J, Hens N, Jit M, Beutels P, Auranen K, et al. (2008) Social contacts and mixing patterns relevant to the spread of infectious diseases. PLoS Med 5: e74.

15. Zagheni E, Billari FC, Manfredi P, Melegaro A, Mossong J, et al. (2008) Using time-use data to parameterize models for the spread of close-contact infectious diseases. Am J Epidemiol 168: 1082-1090.

16. Smieszek T, Burri EU, Scherzinger R, Scholz RW (2012) Collecting close-contact social mixing data with contact diaries: reporting errors and biases. Epidemiol Infect 140: 744-752.

17. Cattuto C, Van den Broeck W, Barrat A, Colizza V, Pinton JF, et al. (2010) Dynamics of person-to-person interactions from distributed RFID sensor networks. PLoS One 5: e11596.

18. Stehle J, Voirin N, Barrat A, Cattuto C, Colizza V, et al. (2011) Simulation of an SEIR infectious disease model on the dynamic contact network of conference attendees. BMC Med 9: 87.

19. Isella L, Stehle J, Barrat A, Cattuto C, Pinton JF, et al. (2010) What's in a crowd? Analysis of face-to-face behavioral networks. J Theor Biol.

20. Salathe M, Kazandjieva M, Lee JW, Levis P, Feldman MW, et al. (2010) A high-resolution human contact network for infectious disease transmission. Proc Natl Acad Sci U S A 107: 22020-22025.

21. Stehle J, Voirin N, Barrat A, Cattuto C, Isella L, et al. (2011) High-resolution measurements of face-to-face contact patterns in a primary school. PLoS One 6: e23176.

22. Hornbeck T, Naylor D, Segre AM, Thomas G, Herman T, et al. (2012) Using Sensor Networks to Study the Effect of Peripatetic Healthcare Workers on the Spread of Hospital-Associated Infections. J Infect Dis.

23. Isella L, Romano M, Barrat A, Cattuto C, Colizza V, et al. (2011) Close encounters in a pediatric ward: measuring face-to-face proximity and mixing patterns with wearable sensors. PLoS One 6: e17144.

24. http://www.sociopatterns.org/

25. Eames KT (2008) Modelling disease spread through random and regular contacts in clustered populations. Theor Popul Biol 73: 104-111.

26. Smieszek T, Fiebig L, Scholz RW (2009) Models of epidemics: when contact repetition and clustering should be included. Theor Biol Med Model 6: 11.

27. Polgreen PM, Tassier TL, Pemmaraju SV, Segre AM (2010) Prioritizing healthcare worker vaccinations on the basis of social network analysis. Infect Control Hosp Epidemiol 31: 893-900.

28. Bernard H, Fischer R, Mikolajczyk RT, Kretzschmar M, Wildner M (2009) Nurses' contacts and potential for infectious disease transmission. Emerg Infect Dis 15: 1438-1444.

29. Lloyd-Smith JO, Schreiber SJ, Kopp PE, Getz WM (2005) Superspreading and the effect of individual variation on disease emergence. Nature 438: 355-359.

30. Temime L, Opatowski L, Pannet Y, Brun-Buisson C, Boelle PY, et al. (2009) Peripatetic health-care workers as potential superspreaders. Proc Natl Acad Sci U S A 106: 18420-18425.

31. (2003) Severe acute respiratory syndrome--Singapore, 2003. MMWR Morb Mortal Wkly Rep 52: 405-411.

32. Botelho-Nevers E, Gautret P, Biellik R, Brouqui P (2012) Nosocomial transmission of measles: an updated review. Vaccine 30: 3996-4001.

33. Voirin N, Barret B, Metzger MH, Vanhems P (2009) Hospital-acquired influenza: a synthesis using the Outbreak Reports and Intervention Studies of Nosocomial Infection (ORION) statement. J Hosp Infect 71: 1-14.

34. Fries J, Segre AM, Thomas G, Herman T, Ellingson K, et al. (2012) Monitoring hand hygiene via human observers: how should we be sampling? Infect Control Hosp Epidemiol 33: 689-695.

35. Abubakar I, Gautret P, Brunette GW, Blumberg L, Johnson D, et al. (2012) Global perspectives for prevention of infectious diseases associated with mass gatherings. Lancet Infect Dis 12: 66-74.

36. Fennelly KP, Martyny JW, Fulton KE, Orme IM, Cave DM, et al. (2004) Cough-generated aerosols of Mycobacterium tuberculosis: a new method to study infectiousness. Am J Respir Crit Care Med 169: 604-609.

37. Halloran ME (2006) Invited commentary: Challenges of using contact data to understand acute respiratory disease transmission. Am J Epidemiol 164: 945-946.

38. Curtis DE, Hlady C, Pemmaraju SV, Polgreen P, Segre AM (2010) Modeling and Estimating the Spatial Distribution of Healthcare Workers. Proceedings of the 1st ACM International Health Informatics Symposium IHI'10: 287-296.

39. Smieszek T, Salathe M (2013) A low-cost method to assess the epidemiological importance of individuals in controlling infectious disease outbreaks. BMC Med 11: 35.

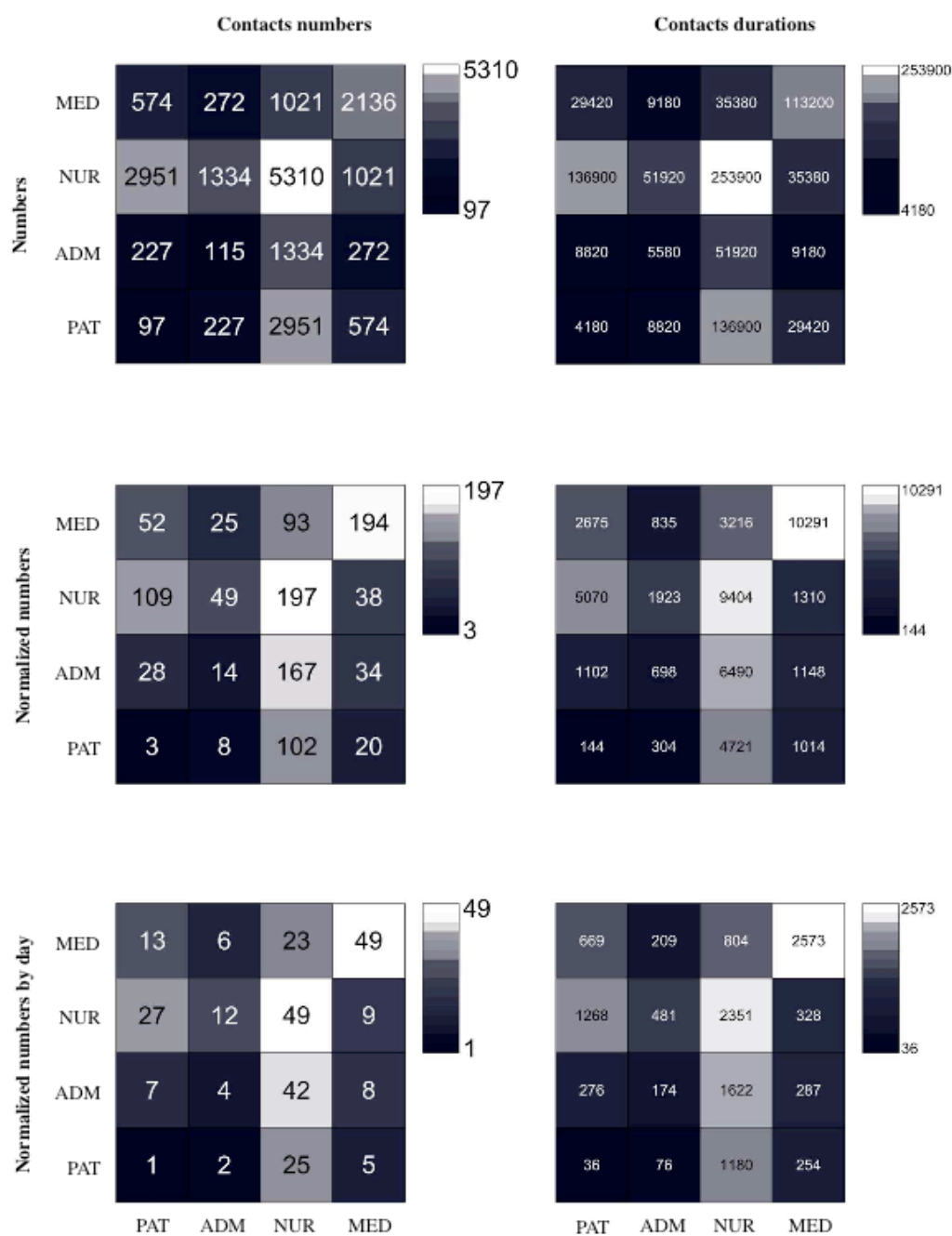

**Figure 1.** Contacts matrices giving the numbers (left column) and cumulated durations in seconds (right column) of the contacts between classes of individuals. In the first line, the matrix entry at row X and column Y gives the total number (resp. duration) of all contacts between all individuals of class with all individuals of class Y. In the second line, the matrix entry at row X and column Y gives the average number (resp. duration) of contacts of an individual of class X with individuals of class Y, during the whole study. In the third line, we normalize each matrix element of the second line matrices by the duration of the study, in days, to obtain average daily numbers and durations of the contacts of an individual of class X with any individual of class Y. The asymmetry of the matrices in the second and third lines is due to the different numbers of individuals populating each class.
Abbreviations: NUR, paramedical staff (nurses and nurses' aides); PAT, Patient; MED, Medical doctor; ADM, administrative staff.

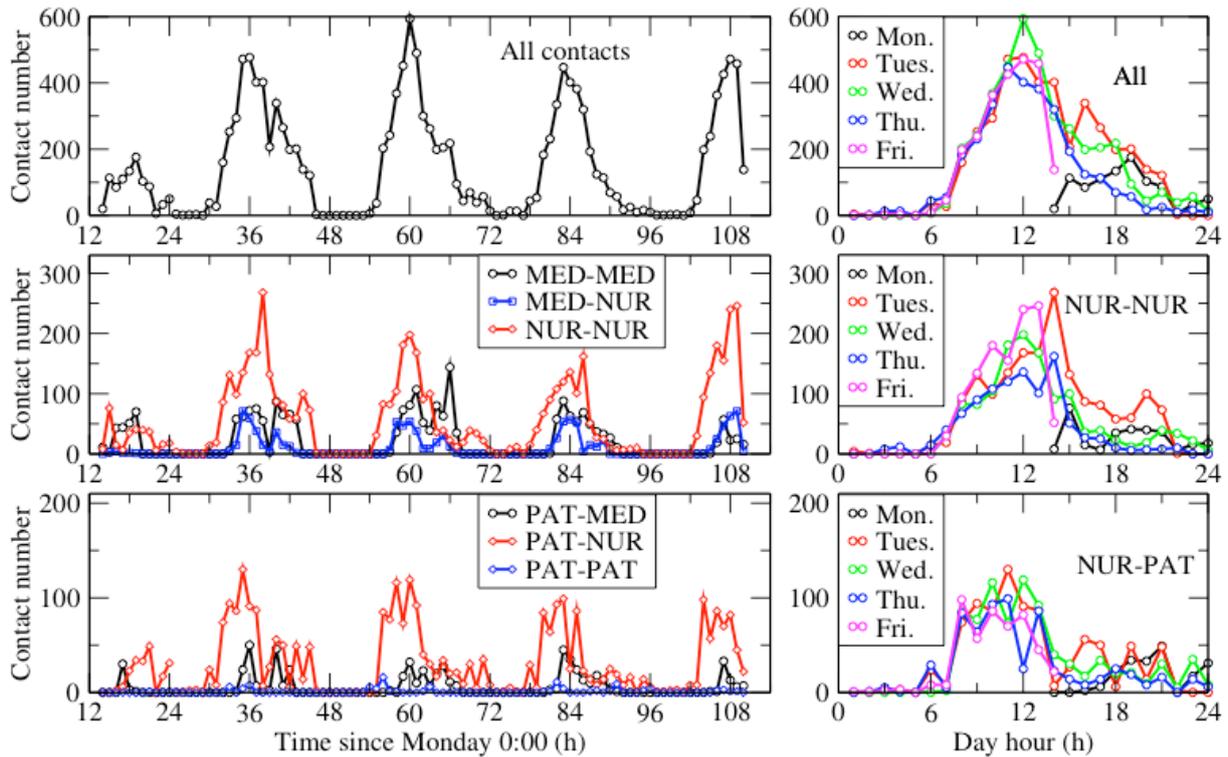

**Figure 2**. Number of contacts per 1-hour periods.
Top row: global number of contacts. Middle and bottom rows: number of contacts involving patients, healthcare workers and medical doctors. The left plots give the number of contacts as a function of the time since the start of the week (Monday, 0:00 AM). The right plots display the number of contacts of several types in each day, as a function of the hour of the day, to show the similarity of the curves in different days.
Abbreviations: NUR, paramedical staff (nurses and nurses' aides); PAT, Patient; MED, Medical doctor; ADM, administrative staff.

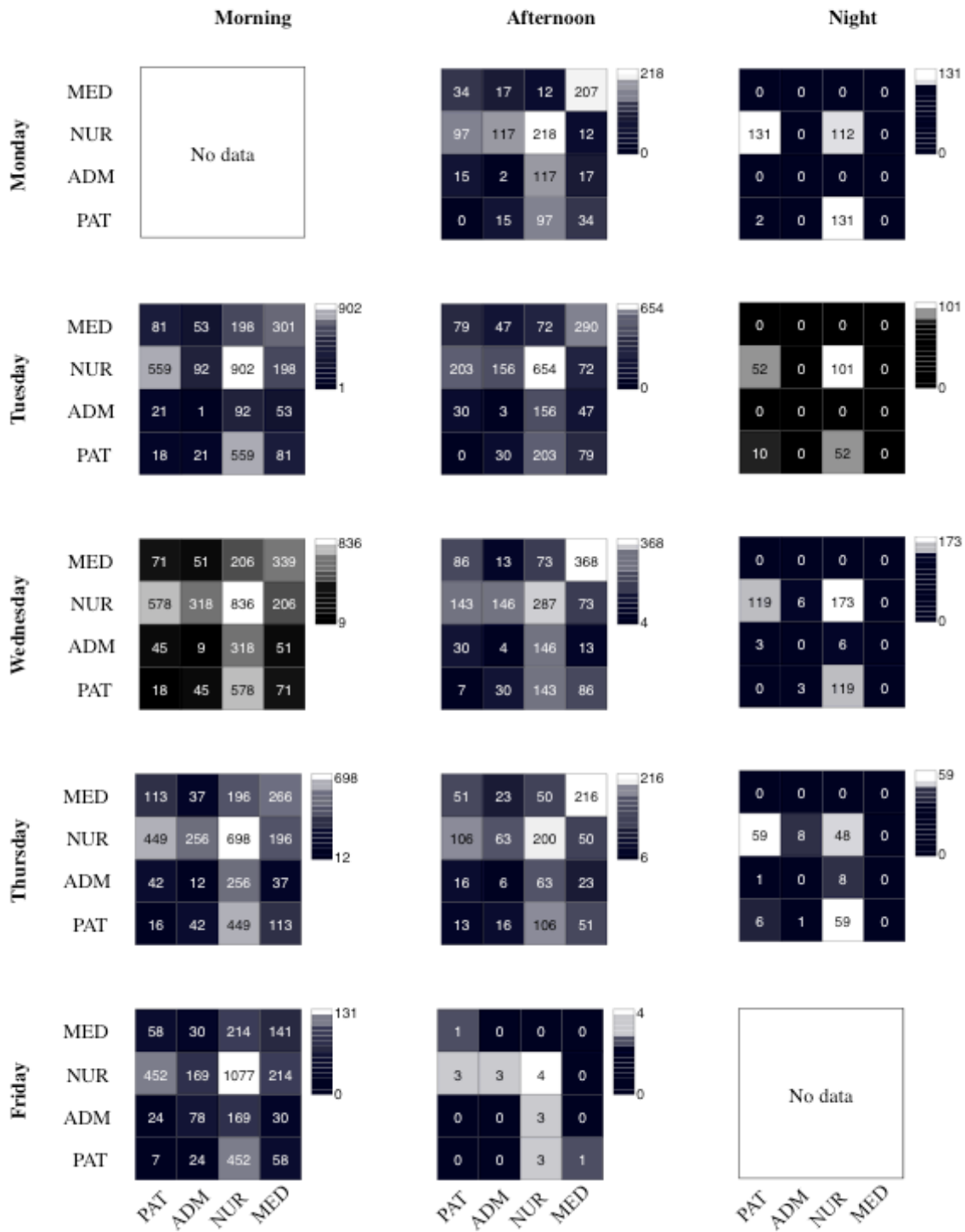

**Figure 3.** Contacts matrices between classes of individuals in each morning, afternoon and night. In each matrix, the entry at row X and column Y gives the total number of contacts of all individuals of class X with all individuals of class Y during each period.

Abbreviations: NUR, paramedical staff (nurses and nurses' aides); PAT, Patient; MED, Medical doctor; ADM, administrative staff.

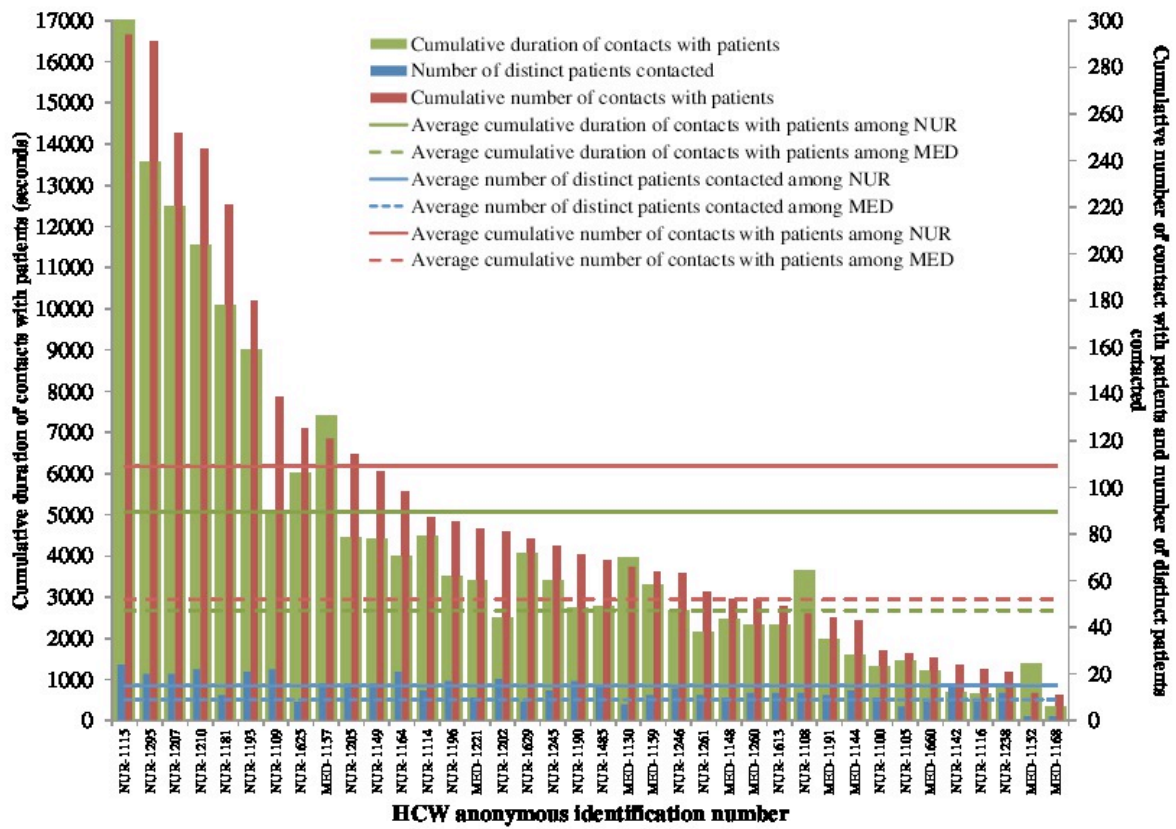

**Figure 4.** Number of distinct patients contacted, number and cumulative duration (in seconds) of contacts with at least one patient for each HCW (NURs and MEDs).

"Super-contactors" are defined as individuals with the highest number of contacts. Here for instance, six HCWs account for more than 40% of the cumulative total of contact numbers and durations.

Abbreviations: HCW, healthcare worker; NUR, paramedical staff (nurses and nurses' aides); MED, Medical doctor.

Table 1. Number of individuals in each class, and average number and duration of contacts during the study per individual in each class. Numbers in parenthesis give the standard deviation.

| Group* | Number of individuals | Average number of contacts per individual (SD) | Average duration (seconds) of contacts per individual (SD) |
|---|---|---|---|
| **NUR** | 27 | 590 (470) | 27111 (24395) |
| **PAT** | 29 | 136 (112) | 6327 (5421) |
| **MED** | 11 | 558 (341) | 27307 (16275) |
| **ADM** | 8 | 258 (291) | 10135 (11439) |
| **Overall** | 75 | 374 (390) | 17293 (19265) |

*Abbreviations: NUR, paramedical staff (nurses and nurses' aides); PAT, Patient; MED, Medical doctor; ADM, administrative staff.

Table 2. Total number and duration of contacts between pairs of individuals belonging to specific classes. Numbers in parenthesis give the percentage with respect to the total number and durations of all detected contacts.

| Pair* | Contact number | Cumulative duration in seconds |
|---|---|---|
| **NUR-NUR** | 5,310 (37.8%) | 253,900 (39.2%) |
| **NUR–PAT** | 2,951 (21.0%) | 136,900 (21.1%) |
| **MED-MED** | 2,136 (15.2%) | 113,200 (17.5%) |
| **NUR–ADM** | 1,334 (9.5%) | 51,920 (8.0%) |
| **MED-NUR** | 1,021 (7.3%) | 35,380 (5.5%) |
| **MED-PAT** | 574 (4.1%) | 29,420 (4.5%) |
| **MED-ADM** | 272 (1.9%) | 9,180 (1.4%) |
| **ADM-PAT** | 227 (1.6%) | 8,820 (1.4%) |
| **ADM-ADM** | 115 (0.8%) | 5,580 (0.9%) |
| **PAT-PAT** | 97 (0.7%) | 4,180 (0.6%) |
| **Total** | 14,037 (100%) | 648,480 (100%) |

*Abbreviations: NUR, paramedical staff (nurses and nurses' aides); PAT, Patient; MED, Medical doctor; ADM, administrative staff.

Table 3. Number and duration of contacts between individuals in the various periods of the days, aggregated over the observation period of 4 workdays and 4 nights.

| | Number of contacts | Cumulative duration of contacts | | |
|---|---|---|---|---|
| | Number (% of total) | Seconds (% of total) | Minutes | Hours |
| **Mornings** | 9,060 (64.5) | 426,860 (65.8) | 7,114 | 118.6 |
| **Afternoons** | 4,165 (29.7) | 185,790 (28.7) | 3,097 | 51.6 |
| **Days** | 13,206* (94.1) | 612,900 (94.5) | 10,215 | 170.3 |
| **Nights** | 831 (5.9) | 35,580 (5.5) | 593 | 9.9 |
| **Total** | 14,037 | 648,480 | 10,808 | 180.1 |

*19 contacts started the morning and ended the afternoon.

**Table 4.** Number and duration of contacts of HCWs (NURs and MEDs) with patients. We show the average of these quantities over all HCWs, all NURs and all MEDs as well as the values for the six HCWs with most contacts.

| RFID tag number and role class | Number of distinct patients contacted | | Cumulative number of contacts with patients | | | Cumulative duration of contacts with patients | | |
|---|---|---|---|---|---|---|---|---|
| | Number | %* | Number | %** | Cumulative %** | Duration | %** | Cumulative %** |
| Average HCW | 13 | 44.8% | 93 | 2.6% | - | 4377 | 2.6% | - |
| Average MED | 9 | 31% | 52 | 1.5% | - | 2675 | 1.6% | - |
| Average NUR | 15 | 51.7% | 109 | 3.1% | - | 5070 | 3% | - |
| 1115 (NUR) | 24 | 82.8% | 294 | 8.3% | 8.3% | 17040 | 10.2% | 10.2% |
| 1295 (NUR) | 20 | 69.0% | 291 | 8.3% | 16.6% | 13580 | 8.2% | 18.4% |
| 1207 (NUR) | 20 | 69.0% | 252 | 7.1% | 23.7% | 12500 | 7.5% | 25.9% |
| 1210 (NUR) | 22 | 75.9% | 245 | 7.0% | 30.7% | 11540 | 6.9% | 32.9% |
| 1181 (NUR) | 11 | 37.9% | 221 | 6.3% | 37.0% | 10080 | 6.1% | 38.9% |
| 1193 (NUR) | 21 | 72.4% | 180 | 5.1% | 42.1% | 9020 | 5.4% | 44.3% |
| Other HCWs | 22 | 75.9% | 139 | 57.9% | 100.0% | 5080 | 55.7% | 100.0% |

*Calculated over the total number of patients in the study, n=29.
** *Calculated over the cumulative number and duration of contacts with patients in the study.